\documentclass{article}
\usepackage[T1]{fontenc} 
\usepackage[utf8]{inputenc} 
\usepackage{ismir,amsmath,cite,url}
\usepackage{graphicx}
\usepackage{color}
\usepackage{tikz}
\usetikzlibrary{decorations.markings}

\usepackage{svg}
\usepackage{adjustbox}
\usepackage{caption}

\usepackage{lineno}
\usepackage{enumitem}

\usepackage[utf8]{inputenc}

\usepackage{float}
\restylefloat{table}

\newcommand{\kdiff}{\scalebox{.85}[1.0]{\tt k-diffusion }}

\usepackage{float}
\usepackage{scalerel}
\usepackage{pgfplots}
\usepackage{ulem}
\usepackage{multirow}
\usepackage{adjustbox}
\usepackage{colortbl}
\usepackage{fancyhdr}
\usepackage{silence}
\usepackage{numprint}
\usepackage{soul}

\usetikzlibrary{datavisualization}
\usetikzlibrary{datavisualization.formats.functions}
\usetikzlibrary{plotmarks}
\usetikzlibrary{angles,quotes,3d,math,arrows.meta,calc,positioning,fit,backgrounds,decorations.pathreplacing,calligraphy,shapes,shapes.multipart}

\let\originalleft\left
\let\originalright\right
\renewcommand{\left}{\mathopen{}\mathclose\bgroup\originalleft}
\renewcommand{\right}{\aftergroup\egroup\originalright}

\newcommand\rurl[1]{%
  \href{https://#1}{\nolinkurl{#1}}%
}

\DeclareMathAlphabet\mathbfcal{OMS}{cmsy}{b}{n}

\pgfplotsset{compat=1.18}

\newcommand{\tikzstylenodedistance}{4mm}
\newcommand{\tikzstyleinnersep}{2mm}
\newcommand{\tikzstyleminimumheight}{7mm}
\newcommand{\tikzstyleminimumwidth}{30mm}

\renewcommand{\tikzstylenodedistance}{3mm}
\renewcommand{\tikzstyleinnersep}{1.25mm}
\renewcommand{\tikzstyleminimumheight}{5.5mm}
\renewcommand{\tikzstyleminimumwidth}{25mm}

\tikzset{
    node distance=\tikzstylenodedistance
}
\tikzset{
    standard node/.style n args={1}{%
        rectangle,
        rounded corners=0.1cm,
        fill=our#1,
        draw=our#1border,
        line width=0.04cm,
        minimum height=\tikzstyleminimumheight,
        minimum width=\tikzstyleminimumwidth,
        inner sep=\tikzstyleinnersep,
        text centered,
        anchor=center,
        align=center,
    }
}
\tikzset{
    standard node circle/.style n args={1}{%
        fill=our#1,
        draw=our#1border,
        circle,
        inner sep=0.1cm,
        minimum height=0,
        minimum width=0,
    }
}
\tikzset{
    standard node circle/.prefix style = standard node
}

\tikzset{
    standard line/.style n args={0}{%
        line width=0.04cm,
        rounded corners=0.1cm,
    }
}
\tikzset{
    standard arrow/.style n args={0}{%
        -latex,
    }
}
\tikzset{
    standard arrow/.prefix style = standard line
}

\definecolor{our}{RGB}{0,0,0}
\definecolor{ourborder}{RGB}{0,0,0}

\definecolor{ourgreen}{RGB}{46, 204, 113}
\definecolor{ourgreenborder}{RGB}{39, 174, 96}
\definecolor{ourblue}{RGB}{52, 152, 219}
\definecolor{ourblueborder}{RGB}{41, 128, 185}
\definecolor{ourorange}{RGB}{230, 126, 34}
\definecolor{ourorangeborder}{RGB}{211, 84, 0}
\definecolor{ourred}{RGB}{231, 76, 60}
\definecolor{ourredborder}{RGB}{192, 57, 43}
\definecolor{ouryellow}{RGB}{241, 196, 15}
\definecolor{ouryellowborder}{RGB}{243, 156, 18}
\definecolor{ourpurple}{RGB}{155, 89, 182}
\definecolor{ourpurpleborder}{RGB}{142, 68, 173}
\definecolor{ourturquoise}{RGB}{26, 188, 156}
\definecolor{ourturquoiseborder}{RGB}{22, 160, 133}
\definecolor{ourturquoise}{RGB}{26, 188, 156}
\definecolor{ourturquoiseborder}{RGB}{22, 160, 133}
\definecolor{ourwhite}{RGB}{236, 240, 241}
\definecolor{ourwhiteborder}{RGB}{189, 195, 199}
\definecolor{ourgray}{RGB}{149, 165, 166}
\definecolor{ourgrayborder}{RGB}{127, 140, 141}

\newif\ifsubmission

\ifsubmission
\linenumbers
\fi 

\title{P\lowercase{ictures} O\lowercase{f} MIDI: 
Controlled Music Generation via Graphical Prompts for Image-Based Diffusion Inpainting}

\ifsubmission
 \threeauthors
  {First Author} {Affiliation1 \\ {\tt author1@ismir.edu}}
  {Second Author} {\bf Retain these fake authors in\\\bf submission to preserve the formatting}
   {Third Author} {Affiliation3 \\ {\tt author3@ismir.edu}}
   \linenumbers
\else
  \oneauthor
 {Scott H. Hawley}
 {Belmont University and Hyperstate AI  
 }
\fi

\ifsubmission
\def\authorname{Anonymous Author}
\else
\def\authorname{S.H. Hawley}
\fi

\usepackage[bookmarks=false,pdfauthor={\authorname},pdfsubject={\papersubject},hidelinks]{hyperref}

\begin{document}

\maketitle
\begin{abstract}
Recent years have witnessed significant progress in generative models for music, featuring diverse architectures that balance output quality, diversity, speed, and user control. This study explores a user-friendly graphical interface enabling the drawing of masked regions for inpainting by an Hourglass Diffusion Transformer (HDiT) model trained on MIDI piano roll images. To enhance note generation in specified areas, masked regions can be ``repainted'' with extra noise. The non-latent HDiT’s linear scaling with pixel count allows efficient generation in pixel space, providing intuitive and interpretable controls such as masking throughout the network and removing the need to operate in compressed latent spaces such as those provided by pretrained autoencoders. We demonstrate that, in addition to inpainting of melodies, accompaniment, and continuations, the use of repainting can help increase note density yielding musical structures closely matching user specifications such as rising, falling, or diverging melody and/or accompaniment, even when these lie outside the typical training data distribution. We achieve performance on par with prior results while operating at longer context windows, with no autoencoder, and can enable complex geometries for inpainting masks, increasing the options for machine-assisted composers to control the generated music. 
\end{abstract}

\section{Introduction}\label{sec:introduction}

Music generation with deep learning models has seen tremendous progress in recent years, enabling algorithmic composition of creative and expressive musical pieces. However, a key challenge that remains is providing users with intuitive control over the generative process. While text prompts and conditioning on attributes like melodies, chords, or timbres have shown promising results \cite{musiclm, borsos2022audiolm, stableaudio}, there is a need to explore richer interactive interfaces that extend user control of constraints on the generative process beyond mere text inputs.

The investigation in this paper is inspired by a typical songwriting composition task in which we seek to create a melody given some accompaniment and a rough idea of when or where we would like the melody to rise and fall \cite{louis}.
We imagine an application in which the user could draw on the piano roll a rough profile for the ``shapes'' of the regions in which to melody is to be added, and have the generative model fill in the notes in way that sounds appropriate, given the accompaniment. This process is somewhat akin to the ``graphic notation'' movement in 20th-century music composition championed by composers such as John Cage \cite{notations}, Cornelius Cardew \cite{Dennis_1991}, and Karlheinz Stockhausen \cite{notations21}, 
by which performers were encouraged to improvise within certain confines indicated by the shapes drawn by the composer.
In this study we imagine not so much the performers improvising, but rather the model generating a composition.
Such visual interfaces for generative models have been present in the image-generation domain for quite some time, such as with NVIDIA's Gau-GAN \cite{gaugan}, and continue today as with the recent ``Semantic Image Synthesis'' work of Park et al\cite{lee2024streammultidiffusion}.

\begin{figure}[b]
    \centering
    \includegraphics[width=.9\columnwidth, trim=9cm 0.7cm 0 1.6cm, clip]{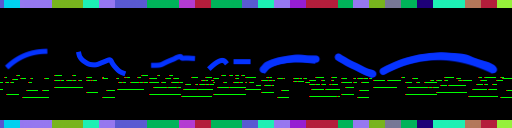}\\
    \includegraphics[width=.9\columnwidth, trim=9cm 0.7cm 0 1.6cm, clip]{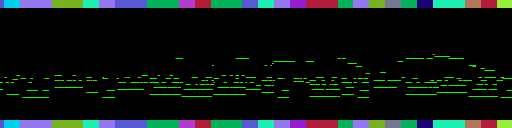}\\
    \caption{The Motivating Idea. Top: MIDI piano roll image of a sample ``graphical prompt'' of rough shapes (in blue) of pitches for melody generation given accompaniment (green lines).  Bottom: Sample generated output.}
    \label{fig:first_prompt}
\end{figure}

An example of such a ``prompt'' is shown graphically in Figure \ref{fig:first_prompt}. One may view the prompt as a constraint on the space of generative possibilities. Such constraints have long been linked with creativity and composition \cite{stravinsky}, as they both limit the field and provide structure around which to generate.   For a task such as this, one may then ask the question which generative model architectures are best suited for this kind of control.  Transformer-based models for musical audio generation such as MusicGen \cite{musicgen} so far have incorporated text prompts and conditioning on melody, however the introduction of arbitrary inpainting masks poses 
 significant challenges for such an architecture. 

Diffusion models have emerged as a powerful framework for generative modeling, achieving state-of-the-art results in domains like image synthesis \cite{stablediffusion, imagen} and audio generation \cite{diffwave, crash, stableaudio}, and even symbolic music generation \cite{mittal2021}. A particularly exciting capability is their support for inpainting or infilling, where partial data is provided as a conditioning context, and the model fills in the missing regions \cite{crash, repaint, eloi_inpaint, chang2021variablelength}. This opens up opportunities for user control via sketching or sculpting of the desired output of the kind shown in Figure \ref{fig:first_prompt}.

 
Image-based control schemes exemplified by ControlNet \cite{controlnet} have inspired musical audio counterparts  \cite{music_controlnet, ditto}. The ``direct'' adaption of image-based music generation was demonstrated notably by Riffusion \cite{riffusion} for audio and Polyffusion \cite{polyffusion} for MIDI. The latter set a new standard for quality, diversity, and controlability of outputs.  MIDI data offers a convenient set of human-readable priors and very compact data representation, making it an excellent testbed for further research in contollable music generation. Polyffusion and others \cite{wang_diffuseroll_2023} explored diffusion models operating on piano roll representations, treating musical scores as image-like data.  While highly effective, these models are typically limited to short sequences due to memory constraints, and may involve the use of compressed latent representations which typically involve basis functions that fail to correspond with humans' intuitive representations.  Our work employs recent advances in high-definition image synthesis with hierarchical diffusion models \cite{hdit, karras2022elucidating} to overcome this limitation and enable longer musical sequences.

Our key interest lies in exploring intuitive interfaces for constrained music infilling, beyond just text prompts. For example, users could sketch a rough target melody profile which the model then turns into a fully-realized composition \cite{tan2022melody}. Inpainting over the compact and human-interpretable piano roll domain allows natural specifications of structural constraints.

The key prior works similar to this study are the aforementioned Polyffusion \cite{polyffusion} for the image-based generative diffusion of piano roll images, and Benetatos et al \cite{Benetatos-2022} offering a drawing-based interface for controllable melody generation. 
Features which this paper lacks are conditioning on chords, rhythm, and/or texture, which have been explored to great effect the aforementioned prior works. Our current codebase does not yet have these features fully implemented. We also focus only on single-instrument generation rather than multi-instrument compositions.  

We believe the following are unique aspects of this paper that contribute to and advance the dialogue on controllable music generation: 
\begin{enumerate}[nosep]
    \item We replicate the results of Polyffusion \cite{polyffusion}, achieving comparable outputs in terms of objective and subjective metrics, yet extending beyond earlier work to include
     \begin{enumerate}[nosep]
         \item (4x) larger images and thus (4x) longer sequences
         \item complex shapes for inpainting masks so users can specify regions and directions of melodic and harmonic development.
         \item explicit modeling of note velocity 
      \end{enumerate}
    \item We apply the recent Hourglass Diffusion Transformer (HDiT) \cite{hdit} that efficiently operates in pixel space on large images, thus 
    \begin{enumerate}[nosep]
        \item removing the need for a separate autoencoder as in \cite{polyffusion}'s use of latent diffusion.
        \item allowing for the straightforward application of complex inpainting mask shapes (that might otherwise need to be remapped into the autoencoder's latent space).
    \end{enumerate}
    
    \item We explore and develop techniques (RePaint \cite{repaint}, nucleation) to encourage note inpainting when outside of training data distribution.
    \item We investigate the limitations of inpainting-only approaches. For example, we find that while inpainting is useful for a variety of tasks, we do not find it to be an effective substitute for chord (progression) conditioning. 
    \item We wish to emphasize that these high quality results have been obtained using the unified, simple, even ``naive'' approach of simply re-appropriating an existing image diffusion codebase, without introducing specialized architectures or mathematical formalism (as in \cite{Benetatos-2022}), 
    and applying inpainting methods commonly seen in image spaces. 
\end{enumerate}

In summary, this paper presents a simple, unified, powerful and flexible framework for controllable and interactive music generation by combining the capabilities of diffusion models, hierarchical representations, and intuitive inpainting constraints specified through a visual interface.

We provide an demo website with listening examples\footnote{Demo website:
\href{https://picturesofmidi.github.io/PicturesOfMIDI}{https://picturesofmidi.github.io/PicturesOfMIDI}\hfill\break 
}

\section{Methods}\label{sec:methods}

\subsection{Data Preparation}\label{subsec:dataset}
Following Polyffusion \cite{polyffusion}, we use the POP909 MIDI dataset \cite{pop909} of piano arrangements of Chinese pop songs. We normalize all tempos to 120 beats per minute to standardize the piano roll representation, and perform data augmentation by transposing up to +/- 12 semitones, expanding the dataset from the initial 909 MIDI files to 22725 files. Piano roll images are rendered such that each pixel corresponds to a 16th note, and as in \cite{polyffusion} we color note durations in green
with brightness corresponding to note velocity, and add red onset markers to add an additional layer of 
verification to rule out spurious diffusion artifacts. 

We apply the same chord detector from the Polyffusion code
to all songs, through which is its found that 529 different chords exist in the augmented (i.e. variably transposed) dataset.  
We encode the chord information as borders along the top and bottom 8 pixels of the images, using a base-9 
arithmetic scheme to separate each color by 30 color-index-values (ranging from 0 to 255) in each R,G,B channel of the image.  The border size of 8 was chosen so as not to overwrite any notes, which given transpositions occur at MIDI pitches as low as 9 and as high as 115.  This leaves the bottom 8 pixels and top 12 pixels available to use for other notations such as embedding chord information.

\begin{figure}
  \centering
  \includegraphics[width=.9\columnwidth]{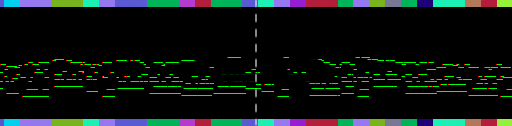}\\
  \vspace{0.1cm}
  \hspace{1.1cm}\begin{tikzpicture}
    \node (image) at (0,0) {\includegraphics[width=.45\columnwidth]{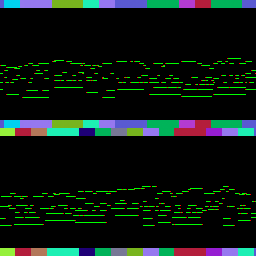}};
    \draw [thick, decoration={markings, mark=at position .9999 with {\arrow{>}}}, postaction={decorate}] (2,1) arc[start angle=90, end angle=-90, x radius=1cm, y radius=1cm];
  \end{tikzpicture}
  \caption{Sample 512x128 MIDI piano roll image. Following Polyffusion \cite{polyffusion}, we denote notes in green with 
  onsets in red. We also add color-coded chord embeddings as borders along the top and bottom of the image. 
    The right half of the image (after the dashed line) is ``folded'' underneath the left half to produce a square  image suitable for example Hourglass Diffusion Transformer (HDiT) \cite{hdit} pipelines.  After generation, the images are 
  restored to their rectangular format. (Although this ``folding'' causes a reversal of direction vs. of a simplw copy-paste, we do this so that information need not propagate all the way across the image to maintain musical continuity. In practice we observe no issues with continuity at fold boundary -- the model quickly learns to adapt.}
\label{fig:pr_1line}
\end{figure}

Each song in each key is rendered as one long $N_t \times 128$ pixel image where $N_t$ is the total number of  16th notes in the song. From these full song images, will train a model by grabbing random 512-pixel windows (padding the ending
with zeros as needed) for each batch of data. Since the HDiT expects square images, we convert each 512x128 
image to 256x256 by putting the second half of the original image below the first. To help the 
network communicate effectively across the cut in the middle, we reverse the lower half of the image horizontally. Figure \ref{fig:pr_1line} shows this process. 
A window width of 512 pixels at 1 pixel per 16th note results in 32 measures, providing approximately 1 minute per image at the standardized tempo of 120 BPM.

\subsection{Neural Network Architecture}\label{subsec:architecture}
We use an image-based deep learning system to learn and generate piano rolls.
Employing image modalities for music representations has a long history in deep learning research, such as the use of spectrogram images for classification \cite{choi}.  The closest musical audio analog of this present study might be Riffusion \cite{riffusion}, which used Stable Diffusion \cite{stablediffusion} essentially unaltered to produce images of spectrograms that were then converted to audio via other means. 

Instead of Stable Diffusion which operates in a latent space defined by a (VQ-)VAE, 
we use the recent Hourglass Diffusion Transformer (HDiT) of Crowson et al \cite{hdit} and the code in the official \kdiff  repository\footnote{\href{https://github.com/crowsonkb/k-diffusion}{https://github.com/crowsonkb/k-diffusion}} with minimal modifications that allow inpainting support.
HDiT operates in pixel space and requires no VAE.
From Crowson's code repository, we minimally modify the example configuration file for the 256x256 pixel Oxford 102 flowers  dataset \cite{oxford_flowers} to our current needs. During training, we
dynamically grab batches of random 512-pixel-wide windows from each $N_t \times 128$ MIDI piano roll image. 
Then, as shown in Figure \ref{fig:pr_1line}, we ``fold'' the right half of each windowed image under the left half to supply a 256x256 image to supply for the HDiT code.

\subsection{Training}\label{subsec:training}
Training was performed for 48 hours on a pair of NVIDIA A6000 GPUs with a batch size of 192. 
Even in as little as the first 1000 training steps, the model is already putting out
discernible notes in green with matching onsets in red, and corresponding chord annotation borders along the top and bottom which mirror each other.  The music
is sheer cacophony at this point, and the remaining two days improve the musicality.
We typically see polynomial convergence of the loss function 
(relative to the number of steps), or slightly worse.  However, eventually
 the model starts memorizing the data, at which point the convergence becomes
 exponential. We halt the training shortly after exponential convergence is first detected.

\subsection{Sampling and Inpainting}\label{subsec:sampling}
Inference for diffusion models is typically referred to as sampling. 
It is during sampling that we do the inpainting described below. The \kdiff code supports a variety 
of higher-order sampling algorithms. We found the Linear Multistep (LMS) sampler to yield poor results when inpainting, and settled on the stochastic variant of the DPM-Solver++(2M) solver \cite{lu2023dpmsolver} instead. (The 3M SDE version of the DPM-Solver++ also produced comparable results.) 


 Important musical inpainting work includes that of CRASH \cite{crash}, the question of achieving usable long-term 
 inpainted music was has received special attention in the literature \cite{eloi_solving, maid_long_inpaint}. Our inpainting
 will rely on a basic binary mask as was outlined in the Polyffusion paper \cite{polyffusion}, although we must adapt the RePaint algorithm \cite{repaint} slightly for the \kdiff package's implementation of the mathematical formalism by Karras et al\cite{karras2022elucidating}. To wit: 
 we compute the $\beta_t$ parameters for Repaint using the $\sigma_t$
 variables of Karras via 
 $$ \beta_t  = ( \sigma_t / \sigma_{\rm max} )^2,$$
where as with comparable HDiT configurations, we use $\sigma_{\rm max}=160$.

RePaint involves repeatedly stepping backwards once re-introducing noise at each stage of the reverse diffusion process, with the number of repaint loops (or ``repaints``) given by some parameter $U$. For generations in Section 2.1 we use $U=1$ (i.e., regular diffusion sampling with no re-paints), but in Section 2.2 we increase $U$ as a way to generate higher note densities in drawn inpainting regions.

Somewhat inspired by the recent work of Lin et al \cite{lin2024arrange} who ``algorithmically render all symbolic controls into the audio format before inputting them to MusicGen,'' we pass chord information into the model as part of each image.
Our hope was to achieve results comparable to those obtained from chord-based conditioning but using inpainting alone instead.  Rather than supply chords as a conditioning signal, we render ``color embeddings'' of the chords and affix them to ``borders'' of unused very high and low pitches, which we observe from our (transposed) dataset to comprise the top and bottom 8 pixels of our 128-pixel-tall MIDI piano roll images.

\section{Results}\label{sec:results}

\subsection{``Standard'' Inpainting Tasks}

\begin{figure}
    \centering
    \includegraphics[width=0.85\columnwidth]{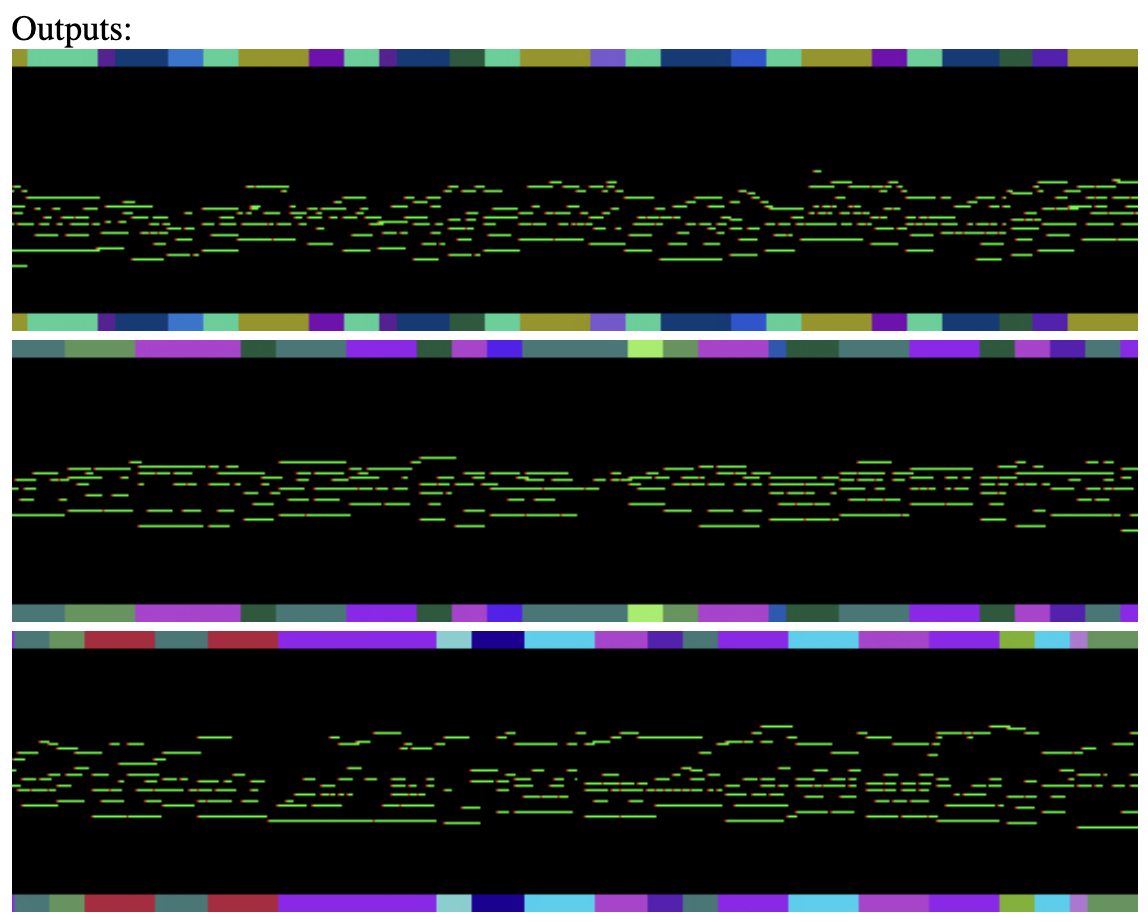}
    \caption{Undirected generation. Here we see model generates various assortments of melody, accompaniment, and ``chord borders''. Refer to the demo
    website for listening examples.  }
    \label{fig:gen-uncond}
\end{figure}

Figure \ref{fig:gen-uncond} shows sample outputs from letting the model generate in an undirectected fashion, based on the learned distribution of training data, with no inpainting.  We see differences in melody, accompaniment, and chords.  Regarding the latter: although the model is able to reliably produce colored chord-marker borders that match on the top and bottom of each image, we find that these colors, when decoded, do not correspond to the chords as detected independently using a chord-extraction code (e.g. from \cite{polyffusion}. This demonstrates a limitation of the ``all-inpainting'' approach in this paper: any chord borders shown in this paper are thus ``decorative'' and may be cropped out. 

\begin{figure}
    \centering
    \includegraphics[width=0.9\columnwidth,  trim=0 5.9cm 0 0, clip]{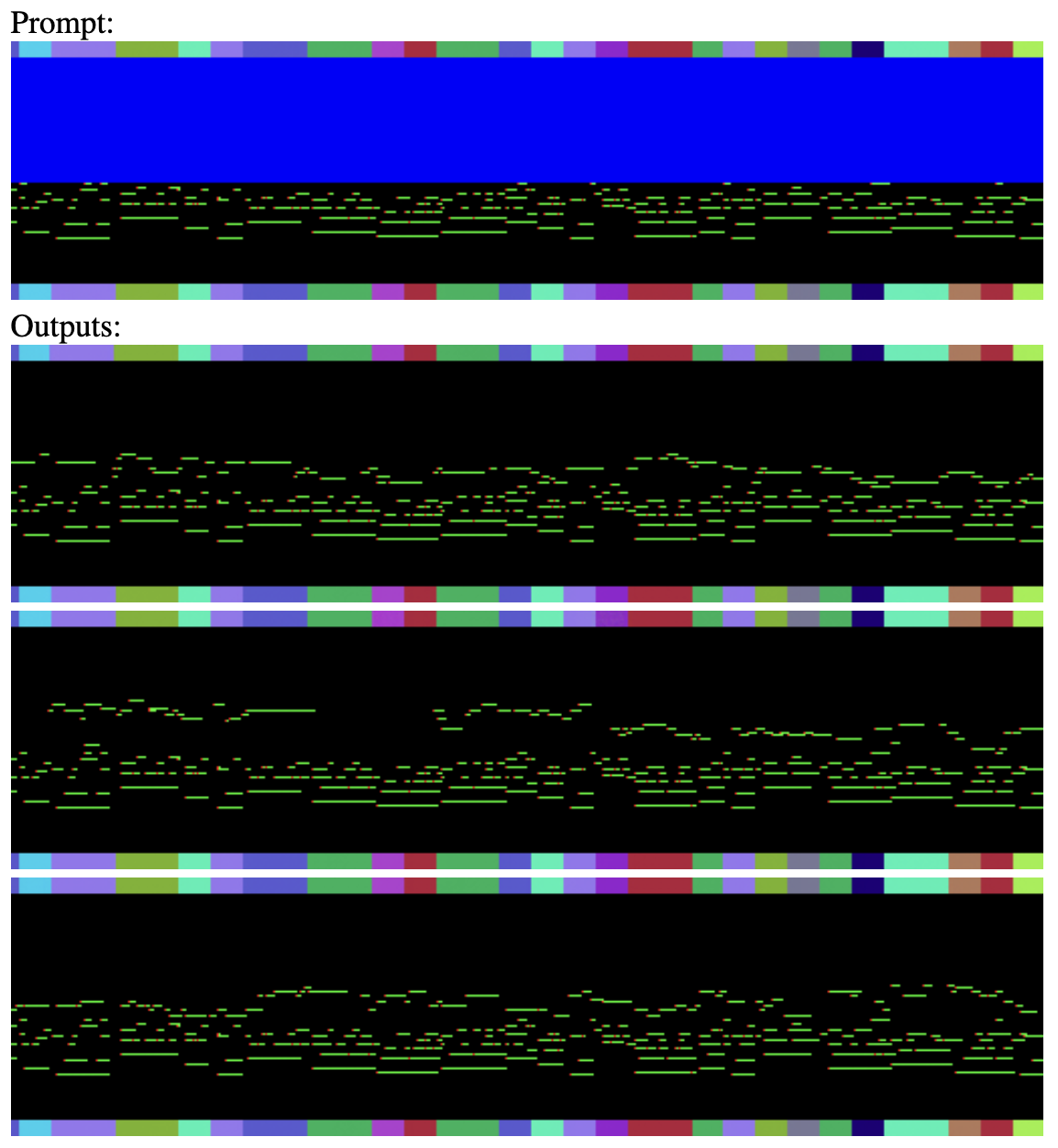}
    \caption{Melody inpainting.  The top portion of the piano roll is masked out in blue (top image), and the model generates melodies (shown in the bottom 3 images) that fit the accompaniment notes and the chords (denoted by colored bars along the top and bottom).}
    \label{fig:gen-melody}
\end{figure}

Figures \ref{fig:gen-melody} through \ref{fig:gen-continue} show examples of common musical inpainting tasks. In Figure \ref{fig:gen-melody}, we mask out the notes in the upper half of the piano roll. The model fills in melody lines consistent with the underlying chord structure and accompaniment.  Figure \ref{fig:gen-accomp} shows the reverse, generating accompaniment given a melody.

\begin{figure}
    \centering
    \includegraphics[width=0.9\columnwidth,  trim=0 5.9cm 0 0, clip]{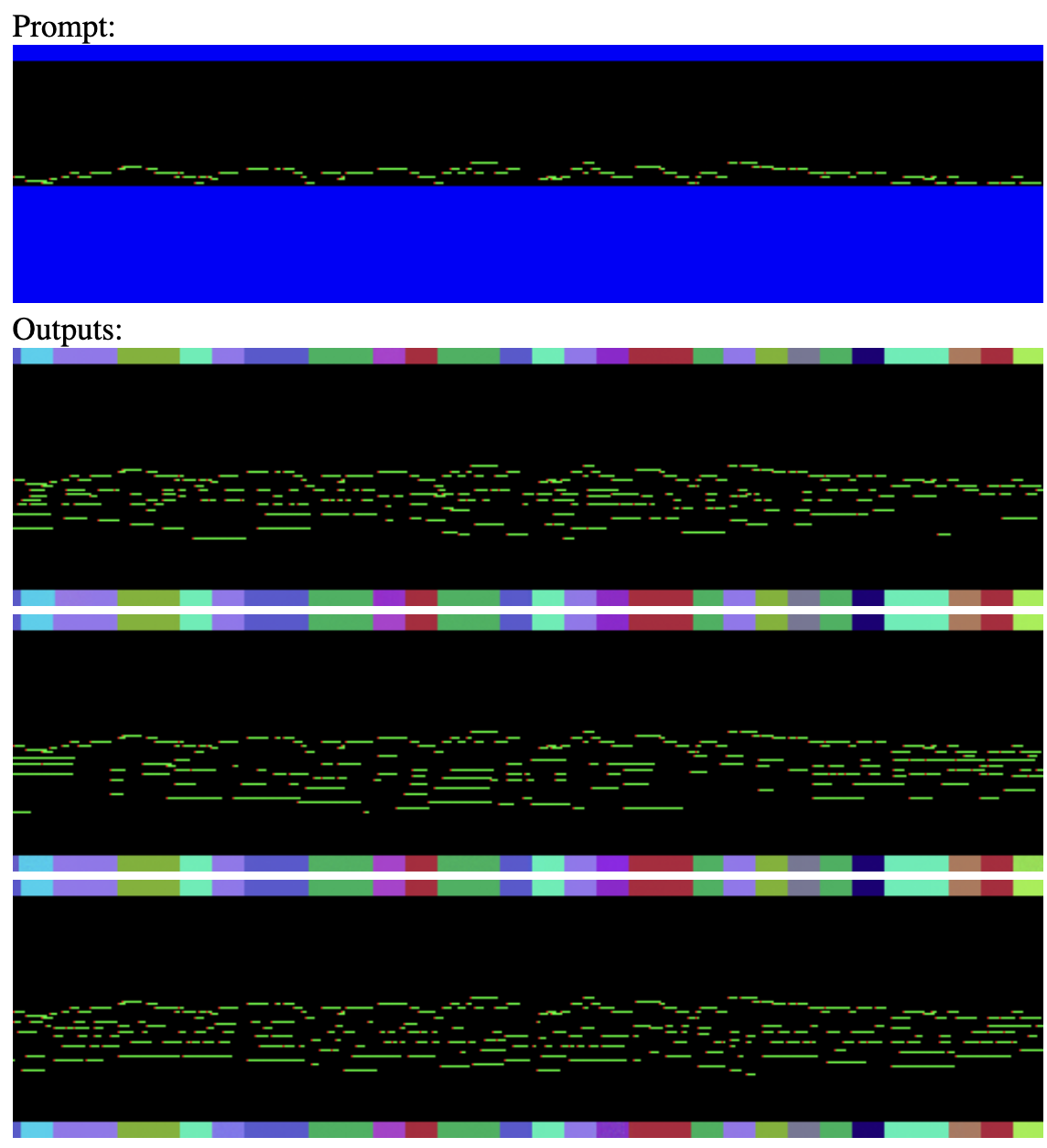}
    \caption{Accompaniment inpainting, given a melody.  It is noteworthy that the chords shown in the top and bottom borders are the same for each output even though the accompaniment notes differ.}
    \label{fig:gen-accomp}
\end{figure}

\begin{figure}
    \centering
    \includegraphics[width=0.9\columnwidth,  trim=0 5.9cm 0 0, clip]{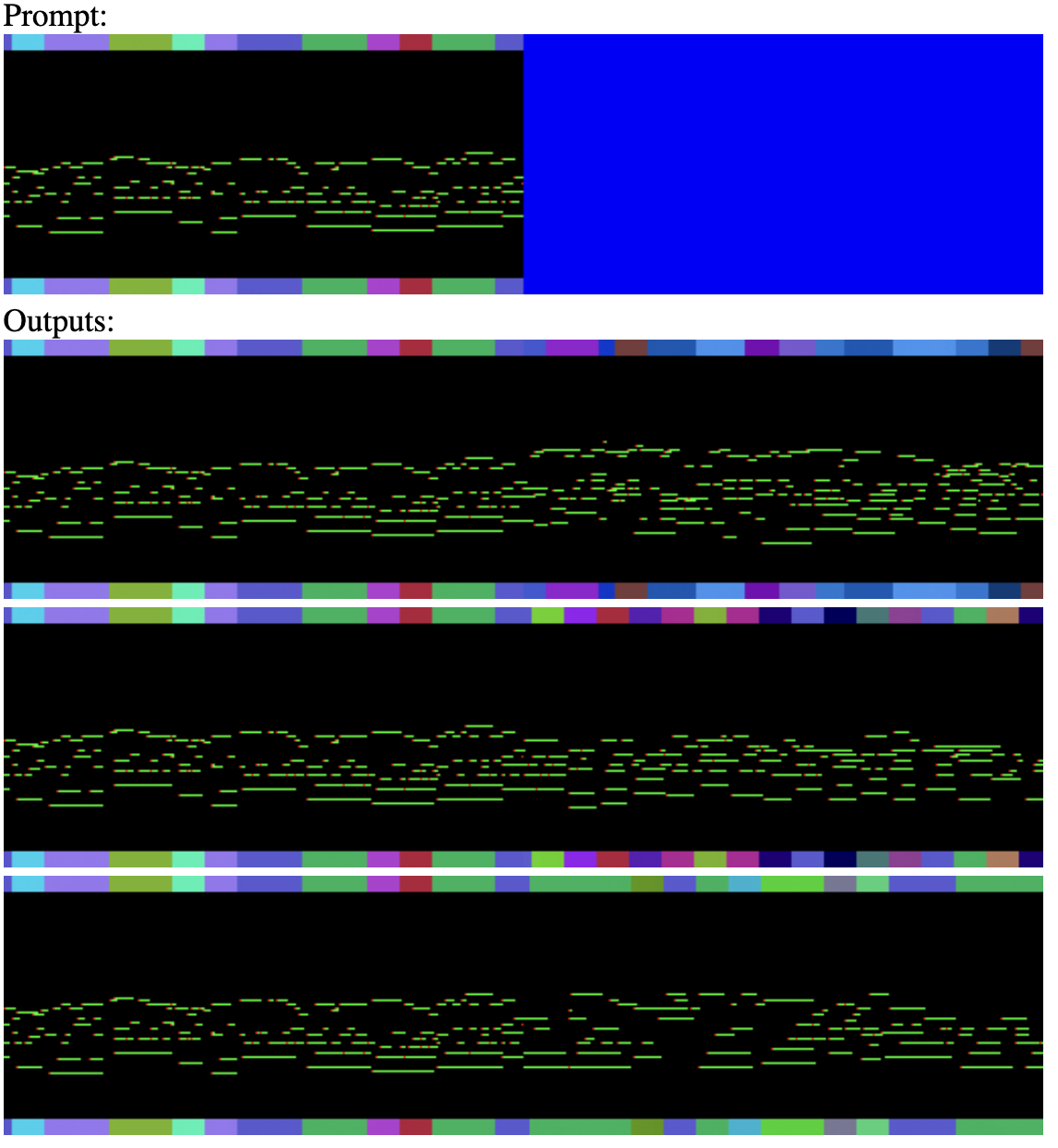}
    \caption{Continuation / "outpainting," offering novel continuations from the same starting music.}
    \label{fig:gen-continue}
\end{figure}

Not shown are the {\it negative} results from 
our inpainting studies: anything involving the chord markers -- inpainting notes given chords and vice versa -- does not work, in the sense that the generated chord markers appear to be arbitrary and do not match the notes at in any significant sense, and the notes generated do not follow the chords requested.  It is possible that a revised inpainting scheme such as RePaint \cite{repaint} would be sufficient to allow these tasks to be completed via inpainting alone, although finding the proper implementation of RePaint for multi-step k-diffusion \cite{karras2022elucidating} integration is thus far an open question.  Otherwise, inpainting alone will need to be abandoned in favor of chord based conditioning, the effectiveness of which was demonstrated by Min et al. \cite{polyffusion}.

\begin{figure}
    \centering 
    \raisebox{0.8cm}{a)}
    \includegraphics[width=0.8\columnwidth]{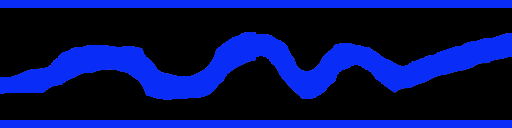}\\
    \raisebox{0.8cm}{b)}
    \includegraphics[width=0.8\columnwidth]{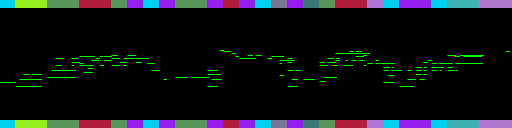}\\

    \caption{Generation constrained by ``creative drawing'' of inpainting mask, Case 1.}
    \label{fig:range-squiggle}
\end{figure}

\subsection{``Creative'' Inpainting Tasks}

We now consider cases where a user may have an idea of the overall shape that they would like the music to follow, expressed via drawing the mask via some graphical interface. (For this paper, these were drawn via GIMP and/or the Mac's Preview app.\footnote{We also have a Gradio GUI demo available on the Demo website.} 
The model is tasked with adding notes that ``make sense'' and meet the user's criteria. 

\begin{figure}
    \centering 
    \raisebox{0.8cm}{a)}
    \includegraphics[width=0.7\columnwidth, trim=0cm 0.7cm 0 1.2cm, clip]{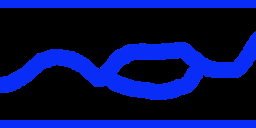}\\
    \raisebox{0.8cm}{b)}
    \includegraphics[width=0.7\columnwidth, trim=0cm 0.7cm 0 1.2cm, clip]{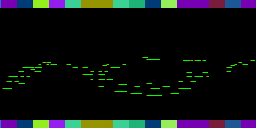}\\
    \caption{Generation constrained by ``creative drawing'' of inpainting mask, Case 3.  Here the user requests that the bass line and melody diverge leaving a sizable gap in between, and then converge.   }
    \label{fig:range-loop}
\end{figure}

The examples shown in Figures \ref{fig:range-squiggle} and \ref{fig:range-loop} are deliberately ``extreme'' in the sense of featuring pitch ranges that fluctuate over larger intervals and shorter timescales than is found in the training data distribution. As a result of this domain mismatch, the model typically produces relatively few notes meeting the user's criteria, instead outputting mostly silence. 

It is possible to ``brute force'' the creation of satisfactory(-looking) outputs by generating a large number of example outputs and ranking them according to the area of the mask filled in by notes (i.e., by taking the dot-product of the mask and the generated image).  These satisfactory outputs are therefore low-probability events. The examples shown in Figures \ref{fig:range-squiggle} and \ref{fig:range-loop} are thus {\it automatically cherry-picked} via this ranking method, obtained by generating 100 outputs at a time and taking only the top 2 outputs in the ranking.  An alternative to this brute force method, namely ``seeding nucleation,'' is explored in the next subsection.

Given that these examples differ significantly from the training data distribution, it is perhaps unsurprising that the generated music is not especially ``musical'' or aesthetically pleasing in the sense of pop tunes. Thus the utility of such extreme masking cases may yet have little practical utility without additional and more varied training data, conditioning signals, and/or diffusion guidance. This remains an area for further study.

\subsection{Increasing Inpainted Note Density}

\begin{figure}
    \centering
    \raisebox{0.8cm}{a)}
    \includegraphics[width=.7\columnwidth, trim=9cm 0.5cm 0 1.5cm, clip]{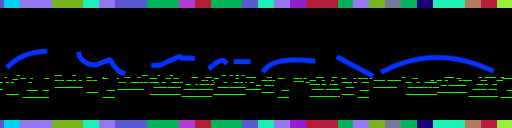}\\
    \vspace*{0.05cm}
    \raisebox{0.8cm}{b)}
        \includegraphics[width=.7\columnwidth, trim=9cm 0.5cm 0 1.5cm, clip]{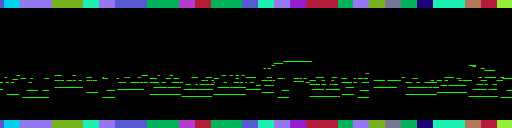}\\
    \vspace*{0.05cm}
    \raisebox{0.8cm}{c)}
    \includegraphics[width=.7\columnwidth, trim=9cm 0.5cm 0 1.5cm, clip]{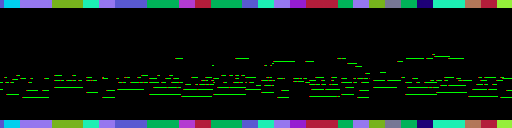}\\
    \vspace*{0.05cm}
    \raisebox{0.8cm}{d)} 
    \includegraphics[width=.7\columnwidth, trim=0cm 0.3cm 0 0.8cm, clip]{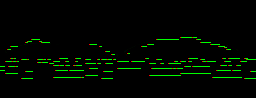}
    \caption{Driving Inpainting.
    a) The user's promp veers a bit high compared to the training data. 
    b) Typical output shows few notes generated.
    c) By brute force (cf. Section 2.2) we can get more notes. 
    d) More efficiently, we can get more notes by increasing the RePaint parameter $U$ from 1 to 2.   As we increase $U$ to 3, 4, or more, we get more notes but they increasingly show little musical coherence and tend toward apparent randomness. 
Our attempts to spell out words in the MIDI in the style of Jacob Collier \cite{collierspells} have so far produced 
nothing close to his level of musicality. 
    }
    \label{fig:draw-melody}
\end{figure}

Depending on how closely or poorly the user's visual prompt conforms to typical data in the training data distribution, the model may have a respectively higher or lower probability of providing matching outputs.
In Figure \ref{fig:draw-melody}a), we see an attempt to generate notes in a mask shape that extends higher than typical 
melodic ranges for POP909.  The results of these generations can be ``hit or miss'' in that there can be a low
probability of notes appearing at these high pitch values as demonstrated in Figure \ref{fig:draw-melody}b).  
Applying the automated brute-forced cherry-picking described in Section 2.2, we can get more notes as shown in Figure \ref{fig:draw-melody}c), however this is computationally needlessly expensive. 
More effective is to combine the ranked cherry-picking method for a smaller number of generated outputs in which we increase the RePaint parameter $U$ from 1 to 2. These results are shown in \ref{fig:draw-melody}d). 

\subsection{Evaluation}

\begin{table*}
 \begin{center}
 \begin{tabular}{l|c|c||c|c|c|c}
  \hline
  Name of Dataset / Model &
  $\sigma_P\uparrow$ & 
  IQR$_D\uparrow$ &
  
  $\mathcal{D}_P\uparrow$ &  $\mathcal{D}_D\uparrow$ & 
  $\mathcal{D}_{\mathrm{KL},P}\downarrow$ &  $\mathcal{D}_{\mathrm{KL},D}\downarrow$  \\
  \hline
Original MIDI                          & 0.46 & 10.6  & n/a & n/a & n/a & n/a    \\
Piano Roll Images $(\pm 12$ semitones) &0.88 & 13.6  & n/a & n/a  & n/a & n/a   \\
\hline
PoM (ours)  
  & 1.0 & 11.7   & 0.87 & 0.95 &  0.036 & 0.14   \\
Polyffusion (our execution) & 0.38 & 11.0  & 0.88 & 0.95  & 0.0050 & 0.13    \\
  \hline
 \end{tabular}
\end{center}
 \caption{Objective Metrics. For MIDI pitches (P) and note durations (D), we measure 
  the diversity of values via the standard deviation of pitches $\sigma_\mathrm{P}$ and the inter-quartile range of durations (IQR$_D$).  
 Similarity to the training data is computed via the average overlapped
area of pitch distribution ( $\mathcal{D}_P$) and duration distribution
( $\mathcal{D}_D$), as well as Kullback–Leibler divergences $\mathcal{D}_\mathrm{KL}$. 
All values should be taken as +/- 1 on the last digit.}
 \label{tab:obj-metrics}
\end{table*}

\begin{figure}
\centering
\includegraphics[width=1.05\columnwidth, trim={2.6cm 0 1cm 0},clip]{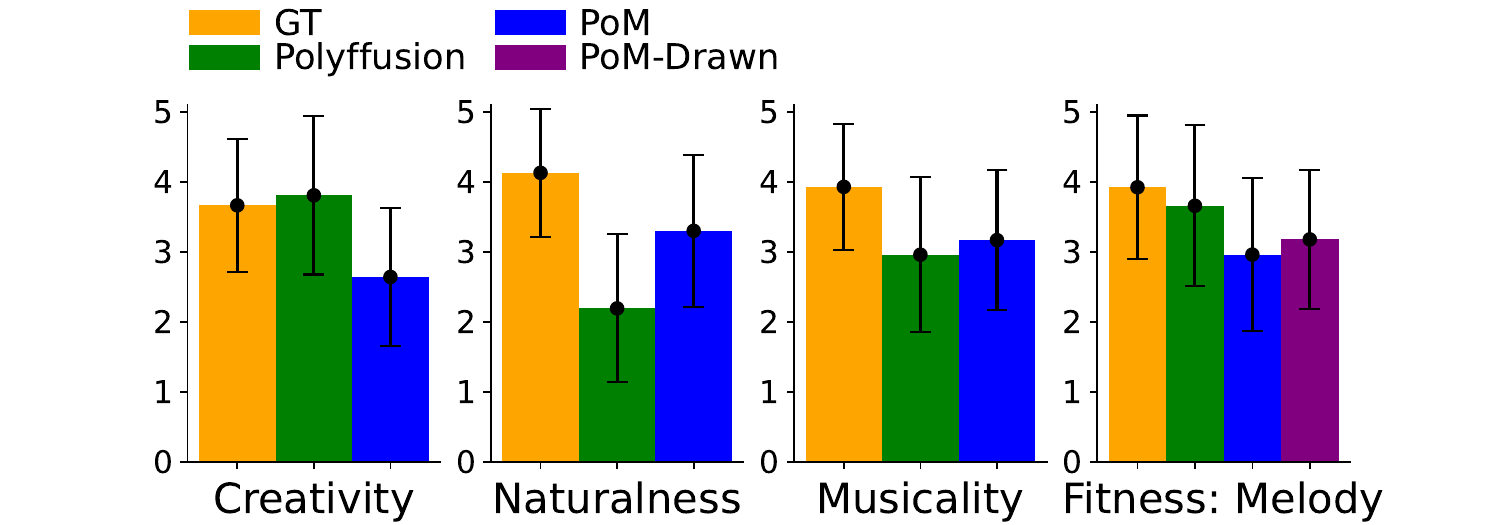}
\centering
\caption{ Subjective evaluation for unconditional generation.  The error bars may suggest that, for the examples provided, listeners found all examples to be comparable. ``GT'' denotes ground truth music from the POP909 MIDI Dataset.  The ``PoM-Drawn'' results are for melodies generated to follow pre-made user-drawn inpainting masks as in Figure \ref{fig:draw-melody}, whereas ``PoM'' alone denotes ``typical'' unconditioned melody generation as in Figure \ref{fig:gen-melody}.}
\label{fig:bar-chart}
\end{figure}

In theory, a desirable model would generate music that is both of high quality and high diversity. While the diversity of outputs can be measured statistically, it is difficult to objectively define ``quality.'' Thus as a proxy for quality, we tend to objectively measure the statistical similarity of the generated data to the training data and supplement this with subjective evaluations using human preferences measured via listening tests. 

For objective diversity measurements, it's appropriate to use the standard deviation of MIDI pitches $\sigma_\mathrm{P}$ since these are approximately normally distributed. The note durations, however, are heavily skewed toward small values, and thus a metric such as the inter-quartile range IQR$_\mathrm{D}$\footnote{This IQR covers 25\% to 75\% of the sorted data points.} is better suited to convey the statistical spread of data rather than the standard deviation (which can be heavily influenced by the the long tail).  Table \ref{tab:obj-metrics} summarizes these measurements for various generative outputs.  For our objective similarity measurements, we compare the training data with generated data by computing the average overlapped area of pitch distribution ($\mathcal{D}_P$ ) and duration distribution ( $\mathcal{D}_D $) as in \cite{polyffusion}. In addition to these we compute Kullback–Leibler divergences $\mathcal{D}_\mathrm{KL}$ of probability density functions (PDF) for each data distribution, obtained via kernel density estimation on histograms of the MIDI pitches ($\mathcal{D}_\mathrm{KL,P}$) and note durations ($\mathcal{D}_\mathrm{KL,D}$).

Subjective listening evaluations were performed by 54 volunteers, half of whom were senior Audio Engineering undergraduate majors, and the remaining half were drawn from the general population. 
We repeated the Polyffusion paper's criteria of ``Creativity,'' ``Naturalness'' (i.e., ``how likely a human musician composed the music''), overall ``Musicality,'' and ``Fitness'' of a generated melody over a fixed accompaniment. 
Evaluators listened to two 16-second long examples drawn from the ``ground truth'' POP909 MIDI dataset, generations from our execution of the Polyffusion code (using their pretrained checkpoints), and our model.  Results are shown in Figure \ref{fig:bar-chart}, which indicate that listeners found all examples to be comparable in quality.  Notably, the ``Fitness'' evaluation included examples of the ``creative inpainting'' variety with a Repaint value of 2 and with drawn melody shapes that, unlike those shown earlier, were of a modest range of notes, thus staying within the typical range of the training data distribution.  
One further finding is that requiring red onset markers for our model makes no difference to our metrics, which  
agrees with the Polyffusion paper \cite{polyffusion}: ``In practice, the generation process
of 160 8-bar samples report zero invalid notes.'' 
Taken together these evaluations suggest that the PoM model generates output comparable to Polyffusion \cite{polyffusion} while supporting longer contexts and arbitrary inpainting mask shapes.

\section{Conclusions}\label{sec:conclusions}

We have presented a unified method for image-based diffusion inpainting that performs comparably to prior inpainting work, yet requires no additional autoencoder due to its use of a Hourglass Diffusion Transformer (HDiT). The HDiT allows for efficient processing of large images allowing longer sequence contexts than previous approaches, and also allows for the straightforward and interpretable use of complex mask geometries for inpainting. These simplifications and extensions, along with the relative ease of appropriating the \kdiff codebase, suggest that image-based piano roll diffusion shows increasing promise for enabling composers to exert 
greater control during the generative music creation process. Future studies may include scaling images up, semi-transparent masks, and multi-instrument piano rolls.

\bibliography{PicturesOfMIDI}

\end{document}